\documentclass{ETHpaper}

\usepackage{amsmath,amssymb,amsfonts}
\usepackage{algorithmic}
\usepackage{graphicx}
\usepackage{textcomp}
\usepackage{xcolor}
\def\BibTeX{{\rm B\kern-.05em{\sc i\kern-.025em b}\kern-.08em
    T\kern-.1667em\lower.7ex\hbox{E}\kern-.125emX}}

\usepackage[numbers]{natbib}
\usepackage{tabularx}
\usepackage{booktabs}
\usepackage[capitalise]{cleveref}
\usepackage{multirow}
\usepackage{nicefrac}
\usepackage{csquotes}
\usepackage{tikz}
\usetikzlibrary{arrows}
\usetikzlibrary{arrows.meta}
\usetikzlibrary{backgrounds}
\usetikzlibrary{calc}
\usetikzlibrary{decorations.pathreplacing}
\usetikzlibrary{fit}
\usetikzlibrary{positioning}

\definecolor{firstcolor}{RGB}{168,50,45} 
\definecolor{colorSG}{RGB}{168,50,45}
\definecolor{customY}{HTML}{FBB13C}
\definecolor{customG}{HTML}{218380}
\definecolor{customT}{HTML}{73D2DE}
\definecolor{customW}{HTML}{FCFCFF}
\definecolor{customB}{HTML}{2E5EAA}
\definecolor{customP}{HTML}{5B4E77}
\definecolor{colorDeveloper}{HTML}{F8766D}
\definecolor{colorDocumenter}{HTML}{A3A500}
\definecolor{colorProductOwner}{HTML}{00BF7D}
\definecolor{colorStakeholder}{HTML}{00B0F6}
\colorlet{colorInactive}{gray}

\colorlet{colorScenarioObserved}{customB}
\colorlet{colorScenarioAgile}{colorSG!40}
\colorlet{colorScenarioRoles}{gray!40}

\newcommand{\code}[1]{\textsc{\detokenize{#1}}}

\newcommand{\redmine}{\textsc{Redmine}}
\newcommand{\aegis}{\textsc{Aegis}}
\newcommand{\git}{\textsc{Git}}
\newcommand{\developer}{\textsc{Developer}}
\newcommand{\developers}{\textsc{Developers}}
\newcommand{\documenter}{\textsc{Documenter}}
\newcommand{\documenters}{\textsc{Documenters}}
\newcommand{\stakeholder}{\textsc{Stakeholder}}
\newcommand{\stakeholders}{\textsc{Stakeholders}}
\newcommand{\productowner}{\textsc{Product Owner}}
\newcommand{\productowners}{\textsc{Product Owners}}

\newcommand{\genua}{\textit{genua}}
\newcommand{\genuaDescription}{German IT security company}
\newcommand{\genuaLong}{\textit{genua GmbH}}

\tikzstyle{legendbullet}=[circle, inner sep=0pt, minimum size=2mm]
\newcommand{\genuaRoleLegend}[1]{
    \begin{tikzpicture}
        \node[legendbullet, fill=colorDeveloper] (bulletDeveloper) at (0, 0) {};
        \node[right=0mm of bulletDeveloper] (labelDeveloper) {\footnotesize \developer{} \ifthenelse{\equal{#1}{1}}{(\code{Dev})}{}};

        \node[legendbullet, fill=colorDocumenter, below=2mm of bulletDeveloper] (bulletDocumenter) {};
        \node[right=0mm of bulletDocumenter] (labelDocumenter) {\footnotesize \documenter{} \ifthenelse{\equal{#1}{1}}{(\code{Doc})}{}};

        \node[legendbullet, fill=colorProductOwner, right=8mm of labelDeveloper] (bulletProductOwner) {};
        \node[right=0mm of bulletProductOwner] (labelProductOwner) {\footnotesize \productowner{} \ifthenelse{\equal{#1}{1}}{(\code{PO})}{}};

        \node[legendbullet, fill=colorStakeholder, below=2mm of bulletProductOwner] (bulletStakeholder) {};
        \node[right=0mm of bulletStakeholder] (labelStakeholder) {\footnotesize \stakeholder{} \ifthenelse{\equal{#1}{1}}{(\code{Sta})}{}};

        \node[anchor=east, left=6mm of $(bulletDeveloper.west)!0.5!(bulletDocumenter.west)$] {\textbf{Role}};
    \end{tikzpicture}
}

\newcommand{\genuastatement}[2]{
    \begin{displayquote}
        \emph{``#1''} #2
    \end{displayquote}
}

\makeatletter
\renewenvironment*{displayquote}
  {\begingroup\setlength{\leftmargini}{.5cm}\csq@getcargs{\csq@bdquote{}{}}}
  {\csq@edquote\endgroup}
\makeatother

\begin{document}

\title{Detecting and Optimising Team Interactions in Software Development}

\makeatletter
\newcommand{\linebreakand}{%
  \end{@IEEEauthorhalign}
  \hfill\mbox{}\par
  \mbox{}\hfill\begin{@IEEEauthorhalign}
}
\makeatother

\author{Christian Zingg\textsuperscript{1,5} \qquad Alexander von Gernler\textsuperscript{2,6} \qquad Carsten Arzig\textsuperscript{2,7} \\ Frank Schweitzer\textsuperscript{1,3,8} \qquad Christoph Gote\textsuperscript{1,4,9}}
\authoralternative{Christian Zingg, Alexander von Gernler, Carsten Arzig, Frank Schweitzer, and Christoph Gote}
\address{
\textsuperscript{1}Chair of Systems Design, ETH Zurich, Zurich, Switzerland \\
\textsuperscript{2}genua GmbH, Kirchheim bei München, Germany\\
\textsuperscript{3}Complexity Science Hub, Vienna, Austria\\
\textsuperscript{4}Data Analytics Group, Department of Informatics, University of Zurich, Zurich, Switzerland \\[2mm]
\textsuperscript{5}czingg@ethz.ch \quad
\textsuperscript{6}alexander\_gernler@genua.de \quad
\textsuperscript{7}carsten\_arzig@genua.de \quad
\textsuperscript{8}fschweitzer@ethz.ch \quad
\textsuperscript{9}cgote@ethz.ch
}

\maketitle

\begin{abstract}
The functional interaction structure of a team captures the preferences with which members of different roles interact.
This paper presents a data-driven approach to detect the functional interaction structure for software development teams from traces team members leave on development platforms during their daily work.
Our approach considers differences in the activity levels of team members and uses a block-constrained configuration model to compute interaction preferences between members of different roles.
We apply our approach in a case study to extract the functional interaction structure of a product team at the \genuaDescription{} \genuaLong{}.
We subsequently validate the accuracy of the detected interaction structure in interviews with five team members.
Finally, we show how our approach enables teams to compare their functional interaction structure against synthetically created benchmark scenarios.
Specifically, we evaluate the level of knowledge diffusion in the team and identify areas where the team can further improve.
Our approach is computationally efficient and can be applied in real time to manage a team's interaction structure.
\end{abstract}

\section{Introduction}\label{sec:introduction}

Designing and maintaining an efficient organisational structure is essential for highly performant software development teams \citep{jassowski2012organizational,tamburri2016architect,gote2021analysing,scholtes2016aristotle,yang2004team,gote2022big}.
This is especially the case in agile software development teams which---similar to Open Source Software teams \citep{Nakakoji2002}---have a strong focus on self-organisation and organisational flexibility \citep{tamburri2022relationship}. 
The key concept behind agile software engineering is a high level of adaptivity, continuous evolution, and flexibility to changes in requirements \citep{beck2001manifesto}.
As a consequence, the real interaction structure of such teams changes over time to adapt to new challenges.
As such, it deviates from the team's original organisational structure to a new \emph{unknown} one \citep{Rank2008,Ralph2016,Valverde2007}.

Not knowing the team's real interaction structure can have a broad range of negative consequences for the team.
In the best case, it leads to reduced productivity or a decrease in software quality due to less well-managed and, therefore, less efficient information exchange \citep{Hoegl2001,Tamburri2015,Bettenburg2010}.
However, in the worst case, it can result in the undetected emergence of developers possessing mission-critical \emph{unshared} knowledge---e.g., \emph{lone wolfs}, \emph{bottlenecks}, or \emph{organisational silos} \citep{Tamburri2015,tamburri2016architect,palomba2018beyond}---which can have a devastating impact when they leave the team \citep{Avelino2016}.

Despite its importance, the question of how to quantitatively and efficiently derive and evaluate a team's functional interaction structure based on real observed interactions remains open.
Closing this gap, we analyse the interactions from a product team at \genuaLong{}, a \genuaDescription{}, and make the following contributions:
\begin{itemize}
    \item Starting from the observed interactions of the team across three development platforms, we show that approaches merely counting the number of interactions between team members fail to detect the real interaction structure, as they cannot account for differences in team members' activities.
    \item Instead, we propose a novel method based on a \emph{block-con\-strai\-ned configuration model} (BCCM) \citep{Casiraghi2019c} that accounts for each team member's unique capacity to initiate and receive (directed) interactions.
    Yielding interaction preferences aggregated on the level of \emph{team member roles}, our method allows us to \emph{quantify} the team's interaction structure on each development platform individually, as well as across all platforms.
    \item We validate the extracted interaction structure through semi-structured interviews with five team members from the product team at \genua{}.
    Using the information obtained from the interviews, we further extend the extracted interaction structure with meta-information on each observed type of interaction.
    As a result, we obtain the team's organigraph \citep{Mintzberg1999}, visualising how different \emph{roles} in a team functionally work together.
    \item Finally, we show how our block model approach can also be used to compare the knowledge diffusion in the observed interactions with two other hypothetical scenarios.
    We find that the team currently achieves knowledge diffusion in the upper third of the possible range.
    Our analysis further shows that extending the agile methods employed by the team is the most promising way to improve knowledge diffusion further.
\end{itemize}

\section{Data}\label{sec:data}

In this paper, we study the case of software development in a product team at \genua{}.
To this end, in \cref{sec:roles}, we first introduce the four roles all members of the product team are subdivided into.
In \cref{sec:interactions}, we then discuss how we mined team interactions from \genua{}'s development platforms.

\subsection{Roles}\label{sec:roles}

Based on their tasks and responsibilities, \genua{} classifies the members of the product team into four roles:
\begin{enumerate}
    \item \textsl{\developers{}} develop, review, and integrate code and changes to implement new features and fix bugs.
    \item \textsl{\documenters{}} write and maintain the user manual and release notes of the product.
    \item \textsl{\productowners{}} coordinate the team and are responsible for scheduling and prioritising issues.
    \item \textsl{\stakeholders{}} 
    only have a peripheral role within the team.
    The majority are customer-facing, selling the product to new clients, maintaining it on their sites, or training their internal staff regarding its use.
    Others perform quality assurance and application testing.
    Finally, some work on other internal projects adjacent to the product.
\end{enumerate}
To obtain the roles for all team members and years, we followed a two-step process.
First, we created lists of all team members active within a given year.
Then, we iterated through these lists with two long-term team members to identify each member's correct role.
In rare cases where the two team members were uncertain regarding a role, they contacted additional team members more familiar with the respective case.
The resulting data set contains (i) team members' IDs and (ii) their roles for (iii) each year.
We provide summary statistics for this data set in \cref{tab:network-role-stats}.

\begin{table}
    \caption{
        Summary statistics on team members and interactions for the four roles.
    }\label{tab:network-role-stats}
    \centering
    \sffamily
    \begin{tabularx}{.7\columnwidth}{X|rr|rr}
        \toprule
                         & \multicolumn{2}{c|}{\textbf{Team members}} & \multicolumn{2}{c}{\textbf{Interactions}}   \\
                         &  Total &  Per year                         &  Total     &  Per year                      \\
        \midrule
        \developers{}     &  67  &  30 -- 51                        &  483,878  &  33,108 -- 73,928               \\
        \documenters{}    &   8  &   3 --  6                        &   19,885  &    180 --  3,614               \\
        \productowner{}  &   5  &   1 --  3                        &   18,888  &    372 --  5,751               \\
        \stakeholders{}   &  62  &  18 -- 40                        &   21,451  &    551 --  5,650               \\
        \bottomrule
    \end{tabularx}
\end{table}

\subsection{Interactions}\label{sec:interactions}

The product team uses three different platforms to track their work.
An \emph{issue tracker} is used to manage and discuss implementations of issues, i.e., bug fixes or feature requests.
When new code to resolve an issue is developed, this is tracked on the team's \emph{code review} platform.
Finally, the team employs \git{} as \emph{version control system} to collaborate on the codebase.
As discussed in the following, for each platform, we mined pseudo-anonymised data capturing all \emph{actions} performed by team members.
In addition, we identified practices to extract the \emph{interactions} corresponding to these actions.
We provide summary statistics for the resulting interactions on all platforms in \cref{tab:network-overview-stats}.

\begin{table}
    \caption{
        Summary statistics on team members and interactions for the three development platforms.
    }\label{tab:network-overview-stats}
    \centering
    \sffamily
    \begin{tabularx}{.7\columnwidth}{X|rr|rr}
        \toprule
                                & \multicolumn{2}{c|}{\textbf{Team members}} & \multicolumn{2}{c}{\textbf{Interactions}}   \\
                                & Total          & Per year                  & Total          & Per year                   \\
        \midrule
        Issue tracker           & 118          & 44 -- 68                 &  77,616       & 3,662 -- 15,745             \\
        Code review platform    &  65          & 17 -- 28                 &  93,256       & 4,759 -- 14,571             \\
        Version control system  &  75          & 34 -- 57                 & 101,179       & 8,152 -- 18,093             \\
        \bottomrule
    \end{tabularx}
\end{table}

\subsubsection{Issue Tracker}

The team uses the tool \redmine{} \citep{Lesyuk2016} as their issue tracker.
Similarly to an online forum, \redmine{} maintains separate discussion threads for all issues.
In \cref{fig:interaction-rules}a we show an example of a discussion thread in which two team members, a stakeholder $S$ and a developer $D$, create entries over time.
The two team members interact when they read each others' discussion entries.
For each discussion entry, we collected (i) the ID of the team member creating it, (ii) the ID of the issue it belongs to, and (iii) the time of the entry's creation.
For reasons of confidentiality, we could not collect the content of the entries.

Together with three members of the product team, we further identified the following two practices ({\sffamily r1}--{\sffamily r2}) that allow us to obtain the interactions corresponding to the creation of each discussion entry:
\begin{enumerate}
    \item[\textbf{\sffamily r1}.] Before team members write their \emph{first} entry in a thread, they read the thread's first entry to read the issue's description.
    Additionally, they read the two most recent messages to learn about the current context of the discussion that their entry will continue.
    \item[\textbf{\sffamily r2}.] For all \emph{subsequent} entries, team members read the thread's first entry to remind themselves of the issue.
    Additionally, they read every entry posted since (and including) their previous discussion entry.
\end{enumerate}
Team member $D$ reading a discussion entry of member $S$ is equivalent to information flowing from $S$ to $D$.
Hence, we model all interactions derived from practices {\sffamily\bfseries r1}--{\sffamily\bfseries r2} as directed links between the author and the reader of a discussion entry.
We illustrate this in \cref{fig:interaction-rules}a, where, for clarity, only the extracted links for $D$ are shown.

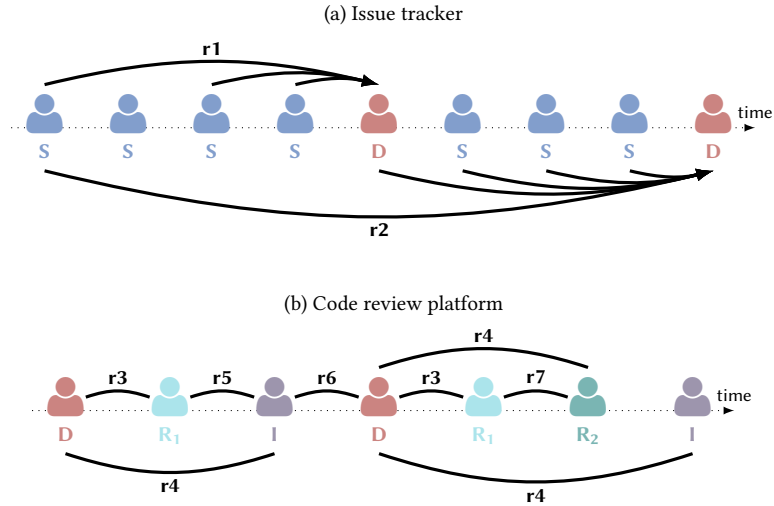
\begin{figure}
    \centering
    \begin{tikzpicture}[xscale=1.1]
        \tikzstyle{mynode} = [font=\sffamily\bfseries\LARGE, align=center]
        \tikzstyle{myedge} = [very thick]
        \tikzstyle{myedgeDir} = [myedge, -latex]
        \tikzstyle{myedgeExt} = [myedge, bend left=20, shorten <= -1mm, shorten >= -1mm]
        \tikzstyle{mylabel} = [font=\sffamily\bfseries\footnotesize, align=center]
        \begin{scope}
            \node[font=\footnotesize] at (4.175,1.5) {(a) Issue tracker};

            \draw[-latex, dotted] (-.4, 0) -- node[above, pos=1] {\scriptsize\sffamily time} (8.5,0);
        
            \node[mynode, customB!60] (n1) at (0,0) {\faUser \\[-4mm] \footnotesize S\vphantom{\textsubscript{1}}};
            \node[mynode, customB!60] (n2) at (1,0) {\faUser \\[-4mm] \footnotesize S\vphantom{\textsubscript{1}}};
            \node[mynode, customB!60] (n3) at (2,0) {\faUser \\[-4mm] \footnotesize S\vphantom{\textsubscript{1}}};
            \node[mynode, customB!60] (n4) at (3,0) {\faUser \\[-4mm] \footnotesize S\vphantom{\textsubscript{1}}};
            \node[mynode, colorSG!60] (n5) at (4,0) {\faUser \\[-4mm] \footnotesize D\vphantom{\textsubscript{1}}};
            \node[mynode, customB!60] (n6) at (5,0) {\faUser \\[-4mm] \footnotesize S\vphantom{\textsubscript{1}}};
            \node[mynode, customB!60] (n7) at (6,0) {\faUser \\[-4mm] \footnotesize S\vphantom{\textsubscript{1}}};
            \node[mynode, customB!60] (n8) at (7,0) {\faUser \\[-4mm] \footnotesize S\vphantom{\textsubscript{1}}};
            \node[mynode, colorSG!60] (n9) at (8,0) {\faUser \\[-4mm] \footnotesize D\vphantom{\textsubscript{1}}};

            \draw[black] (n1.north) edge[myedgeDir, bend left=15] node[above=-.5mm, mylabel] {r1} (n5.north);
            \draw[black] (n3.north) edge[myedgeDir, bend left=15] (n5.north);
            \draw[black] (n4.north) edge[myedgeDir, bend left=15] (n5.north);
            \draw[black] (n1.south) edge[myedgeDir, bend right=15] node[below=-.5mm, mylabel] {r2} (n9.south);
            \draw[black] (n5.south) edge[myedgeDir, bend right=15] (n9.south);
            \draw[black] (n6.south) edge[myedgeDir, bend right=15] (n9.south);
            \draw[black] (n7.south) edge[myedgeDir, bend right=15] (n9.south);
            \draw[black] (n8.south) edge[myedgeDir, bend right=15] (n9.south);
        \end{scope}

        \begin{scope}[yshift=-3.75cm, xshift=.25cm]
            \node[font=\footnotesize] at (3.925,1.4) {(b) Code review platform};

            \draw[-latex, dotted] (-.4, 0) -- node[above, pos=1] {\scriptsize\sffamily time} (8,0);
        
            \node[mynode, colorSG!60] (m1) at (0,0) {\faUser \\[-4mm] \footnotesize D\vphantom{\textsubscript{1}}};
            \node[mynode, customT!60] (m2) at (1.25,0) {\faUser \\[-4mm] \footnotesize R\textsubscript{1}};
            \node[mynode, customP!60] (m3) at (2.5,0) {\faUser \\[-4mm] \footnotesize I\vphantom{\textsubscript{1}}};
            \node[mynode, colorSG!60] (m4) at (3.75,0) {\faUser \\[-4mm] \footnotesize D\vphantom{\textsubscript{1}}};
            \node[mynode, customT!60] (m5) at (5,0) {\faUser \\[-4mm] \footnotesize \footnotesize R\textsubscript{1}};
            \node[mynode, customG!60] (m6) at (6.25,0) {\faUser \\[-4mm] \footnotesize \footnotesize R\textsubscript{2}};
            \node[mynode, customP!60] (m7) at (7.5,0) {\faUser \\[-4mm] \footnotesize I\vphantom{\textsubscript{1}}};

            \draw[black] (m1.30) edge[myedgeExt] node[above=-.5mm, mylabel] {r3} (m2.150);
            \draw[black] (m2.30) edge[myedgeExt] node[above=-.5mm, mylabel] {r5} (m3.150);
            \draw[black] (m3.30) edge[myedgeExt] node[above=-.5mm, mylabel] {r6} (m4.150);
            \draw[black] (m4.30) edge[myedgeExt] node[above=-.5mm, mylabel] {r3} (m5.150);
            \draw[black] (m5.30) edge[myedgeExt] node[above=-.5mm, mylabel] {r7} (m6.150);

            \draw[black] (m1.south) edge[myedge, bend right=20] node[below=-.5mm, mylabel] {r4} (m3.south);
            \draw[black] (m4.south) edge[myedge, bend right=20] node[below=-.5mm, mylabel] {r4} (m7.south);
            \draw[black] (m4.north) edge[myedge, bend left=20] node[above=-.5mm, mylabel] {r4} (m6.north);
        \end{scope}
    \end{tikzpicture}
    \caption{
        Visualisation of the identified practices {\sffamily\bfseries r1}--{\sffamily\bfseries r7} to derive interactions.
        For descriptions of the practices, we refer to the text.
        (a) Directed interactions derived from an exemplary discussion thread between two team members $S$ and $D$ on the issue tracker. For clarity, we only show the interactions derived for $D$.
        (b) Undirected interactions derived from an exemplary change development on the code review platform.
        On this platform, team members develop $D$, review $R$, or integrate $I$ changes. In the example, the change is reviewed by two separate team members, $R_1$ and $R_2$. As indicated by the colours, the developer, reviewer, and integrator must be different team members.
    }\label{fig:interaction-rules}
\end{figure}

\subsubsection{Code Review Platform}\label{sec:aegis}

To resolve issues, team members need to develop, review, and ultimately integrate \emph{changes} to the codebase.
This process is tracked and managed on the code review platform \aegis{} \citep{Miller2013}.
For \aegis{}, we again mined all actions of team members related to all changes.
Specifically, we extracted (i) the ID of the team member performing an action, (ii) the ID of the corresponding change, (iii) the time at which the action was performed, and (iv) the type of the action.
The possible types of actions are development $D$, review $R$, and integration $I$.
The developer, reviewer, and integrator of a change must be different team members.
In the ideal case, a change is first developed, then positively reviewed, and finally successfully integrated.
However, both review and integration can fail, requiring further development and, hence, resulting in more complex action sequences (see \cref{fig:interaction-rules}b for an example).

The change development process requires extensive interactions between team members that can be derived from the recorded actions following practices {\sffamily\bfseries r3}--{\sffamily\bfseries r7}, visualised in \cref{fig:interaction-rules}b:
\begin{enumerate}
    \item[\textbf{\sffamily r3}.]
    A reviewer $R$ discusses the review's outcome with the developer $D$ of the change.
    \item[\textbf{\sffamily r4}.]
    An integrator $I$  discusses the integration's outcome with the developer $D$ of the change.
    \item[\textbf{\sffamily r5}.]
    If the integration fails, the integrator $I$ further discusses the detected problems with the reviewer $R$ that positively reviewed the change.
    \item[\textbf{\sffamily r6}.]
    A developer $D$ that continues development after a failed review or integration discusses with the corresponding previous reviewer $R$ or integrator $I$.
    \item[\textbf{\sffamily r7}.]
    If a developer $D_2$, reviewer $R_2$, or integrator $I_2$ take over from a previous developer $D_1$, reviewer $R_1$, or integrator $I_1$, a handover discussion takes place.
\end{enumerate}
As all discussions resulting from {\sffamily\bfseries r3}--{\sffamily\bfseries r7} are bi-directional, we model them as undirected links between the involved team members.

\subsubsection{Version Control System}

From the \git{}-based version control system, we obtain interactions by extracting yearly co-editing networks using the Python package \texttt{git2net}~\citep{Gote2019}.
Motivated by the finding that a significant proportion of coordination between developers occurs via the code base \citep{Bolici2016}, specifically when editing the same code \citep{cataldo2006identification}, co-editing networks link team members consecutively modifying the same line of code.
The links are directed following the arrow of time and time-stamped according to the time of the edit.
As for the other two platforms, team members are represented by their pseudo-anonymised ID.

\section{Interaction Structure Detection}\label{sec:quantitative_analysis}

Having introduced \genua{}'s product team, how can we detect its functional interaction structure, i.e., the interaction preferences of the different roles?
As an intuitive approach, we could simply count the interactions between the different roles.
In \cref{sec:simple-approach}, we show the results of this approach.
We then discuss its shortcomings in \cref{sec:shortcomings-counting} and propose a more comprehensive block model approach allowing us to overcome them in \cref{sec:bccm-approach}.
Finally, in \cref{sec:bccm-application-to-genua}, we use this approach to detect the functional interaction structure of the studied product team at \genua{}.

\subsection{Interaction counting approach}\label{sec:simple-approach}

The functional interaction structure of a team describes the interaction preference members of each role have towards the members of other roles.
Therefore, it appears natural to simply count for each pair of roles how often their members interact and then identify which pairs have many interactions.
In \cref{fig:count-networks}a, we show these counts for the different development platforms used by the product team at \genua{}.
On each platform, there is one role whose members are inactive.
\documenters{} do not use the issue tracker but instead track the changes in the documentation entirely via the code review platform.
Similarly, \stakeholders{} interact on the issue tracker but do not appear on the code review and version control platforms.
For all platforms, we observe interactions between all active roles.
As indicated by the width of links, the interaction counts are in the same order of magnitude and low compared to the number of interactions observed among \developers{}.
These conclusions also hold when combining the interactions from the three platforms.
As we see in \cref{fig:count-networks}b, \developers{} frequently interact among themselves, and the only missing interactions are between \stakeholders{} and \documenters{} as they are not active on the same platform.

\begin{figure*}
    \includegraphics[width=\textwidth]{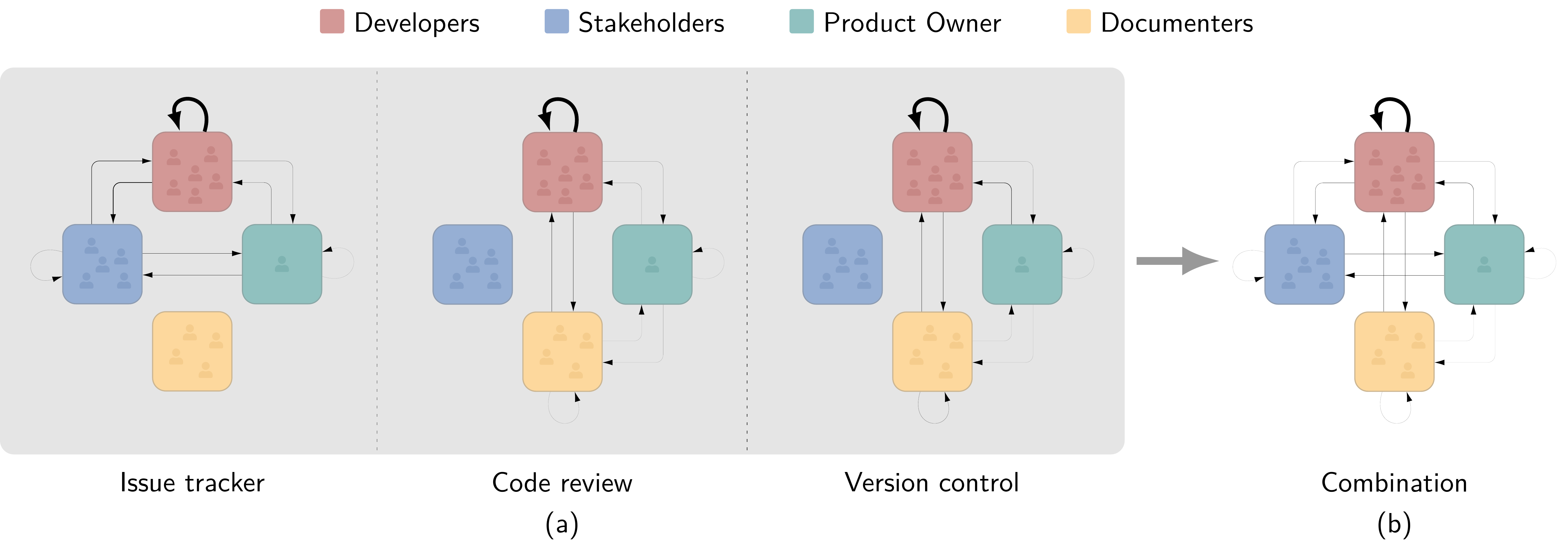}
    \caption{
        Interaction structure extracted using the interaction counting approach. Nodes represent the four roles, and the width of links is proportional to the respective number of interactions between 2010 and 2018.
        The link widths indicate that except for interactions among \developers{}, all interaction counts are low and in the same order of magnitude.
        (a) Interaction structure for each platform. (b) Combined interaction structure for the three platforms.
    }\label{fig:count-networks}
\end{figure*}

\subsection{From interaction counts to interaction preferences}\label{sec:shortcomings-counting}

The interaction counting approach only shows the activity of each role and disregards:
\begin{enumerate}
    \item[(i)] how many interactions each role \emph{can} initiate, and
    \item[(ii)] how many interactions each role \emph{can} receive.
\end{enumerate}
To highlight the consequences of this, we consider the synthetic example with 6 individuals from 3 roles shown in \cref{fig:pot-explanation}a.
The example assumes 2 \stakeholders{} interacting among themselves with moderate activity and 3 \developers{} interacting among themselves but with high activity.
In addition, 1 \productowner{} coordinates between the two groups.
This \productowner{} interacts 50 times with \developers{} but only 10 times with \stakeholders{}.
Does this mean that the \productowner{} has an interaction preference with \developers{}?
While this seems to be the case according to the interaction counting approach, we argue that it is not.
\developers{} appear in 1550 interactions in total, whereas \stakeholders{} appear in \emph{only} 30 interactions.
The \productowner{} is involved in 10 of these 30 interactions, which suddenly seems like a lot.

\subsection{Block model approach}\label{sec:bccm-approach}

\begin{figure*}[t!]
    \includegraphics[width=\linewidth]{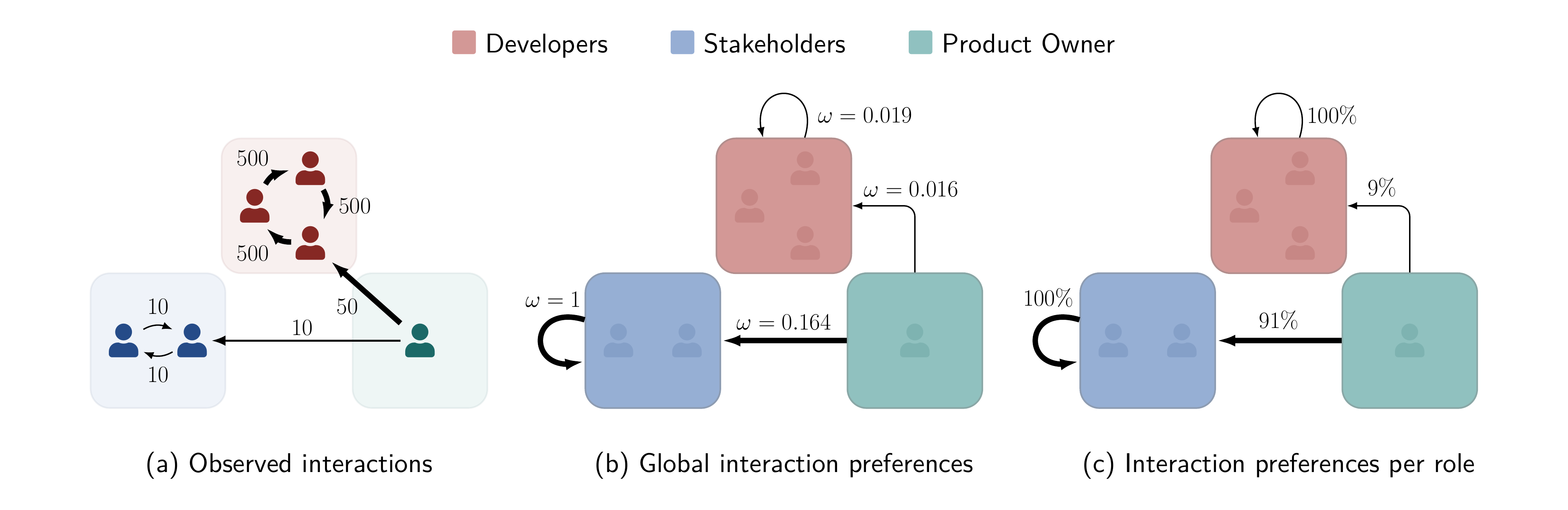}
    \caption{
        Applying our block model approach to extract the functional interaction structure in a synthetic example. (a) Counts of observed interactions between six team members belonging to three roles. (b) Global interaction propensities calculated for these observation counts. (c) Normalised interaction propensities for each role.
    }\label{fig:pot-explanation}
\end{figure*}

We control for the activity of roles in our computation of the interaction structure using a \emph{block-con\-strai\-ned configuration model}~(BCCM) \citep{Casiraghi2019c}, in which we define the blocks as the roles of the team members.
We select the BCCM over the more established stochastic block model (SBM), as the SBM cannot correctly reproduce the empirically observed activities of the team members \citep{Casiraghi2019c}.
To fit the BCCM to observed interactions, these need to be given as an \emph{interaction network} where the nodes are the team members and the (multi-)edges their (multiple) interactions.
For an example, we refer to \cref{fig:pot-explanation}a.
The BCCM introduces a term $\omega_{r_1 r_2}$ that measures the \emph{propensity} with which the roles $r_1$ and $r_2$ interact according to their in- and out-degrees.
To determine $\omega_{r_1 r_2}$ from a given interaction network, we use the function \code{bccm} in the R library \code{ghypernet}~\citep{ghypernet}, which implements:
\begin{equation}\label{eq:bccm-omega}
    \omega_{r_1r_2} := -\log{\left(1-\frac{A_{r_1r_2}}{\Xi_{r_1r_2}}\right)}.
\end{equation}
Here, $r_1$ and $r_2$ are two roles, $A_{r_1r_2}$ is the number of interactions between team members with these roles, and $\Xi_{r_1r_2}$ is a normalisation term for the maximum number of possible interactions between $r_1$ and $r_2$.
For a detailed discussion of \cref{eq:bccm-omega} we refer to \citep{Casiraghi2019c}.

\cref{fig:pot-explanation}b shows the $\omega$ values between all pairs of roles for our example with the interaction counts shown in \cref{fig:pot-explanation}a.
Despite the relatively low interaction count of 10 between the \productowner{} and \stakeholders{}, the low activity of \stakeholders{} results in a comparatively large $\omega = 0.164$ between the two roles.
In contrast, the high activity of \developers{} yields a ten-times lower $\omega = 0.016$ between \productowner{} and \developers{}, despite 50 interactions taking place.
Thus, the BCCM suggests that the \productowner{} is ten times as likely to interact with \stakeholders{} than with \developers{} when taking into account that \stakeholders{} are less active than \developers{}.
This is also reflected in \cref{fig:pot-explanation}c, where we normalise the interaction propensities $\omega$ for each role individually.

If we have no interaction preferences, we would expect equal normalised interaction propensities of $33\%$\footnote{Or $50\%$ if we disregard possible interactions among \productowners{} as there usually is just one.} towards all three roles.
The deviation from this expectation correctly reveals a strong \emph{positive interaction preference} of the \productowner{} towards \stakeholders{} and a strong \emph{negative interaction preference} towards \developers{}.

\subsection{Detecting \genua{}'s interaction structure}\label{sec:bccm-application-to-genua}

As shown in \cref{fig:bias-networks}, we now apply our block model approach to detect the functional interaction structure of the studied product team at \genua{}.
Based on the interaction networks for multiple platforms and multiple years that we collected in \cref{sec:data}, we first compute the interaction preferences separately for each platform and year before averaging them across the years.
As our team contains members of four roles, we consider the threshold to distinguish between positive and negative preferences to be $25\%$\footnote{The threshold of $25\%$ assumes that interactions can occur between members of the same role. This is intuitively true for \developers{}, \stakeholders{}, and \documenters{} as there are always multiple active members of these roles. The \productowner{} is a special case. While in principle, there is only one \productowner{} active at any point in time, our data contains multiple transitions between \productowners{}, resulting in two \productowners{} being recorded for a year. Therefore, we opted to treat \productowners{} analogous to the other roles and did not introduce an exception.}.
In \cref{fig:bias-networks}b, we show the resulting positive and negative interaction preferences.

\paragraph{Positive interaction preferences}
On the issue tracker, \stakeholders{}, the \productowner{}, and \developers{} are active.
The interaction preferences suggest that \stakeholders{} predominantly interact with themselves and the \productowner{}.
In turn, the \productowner{} has an interaction preference towards \developers{}.
Finally, \developers{} again preferentially interact among themselves.\\
The code review platform has activity from all roles other than \stakeholders{}.
Here, the interaction preferences suggest interactions from the \productowner{} towards both \developers{} and \documenters{}, who in turn show a preference to interact among themselves.\\
Finally, the version control system shows a similar pattern as the code review platform.
Again, only the \productowner{}, \developers{}, and \documenters{} are active.
The \productowner{} has an interaction preference towards \developers{}, and \developers{} and \documenters{} interact primarily among themselves.
The only difference to the code review platform is the absence of a positive preference between the \productowner{} and \documenters{}.

\begin{figure*}
    \centering
    \includegraphics[width=\textwidth, trim=4cm 0cm 0cm 0cm, clip]{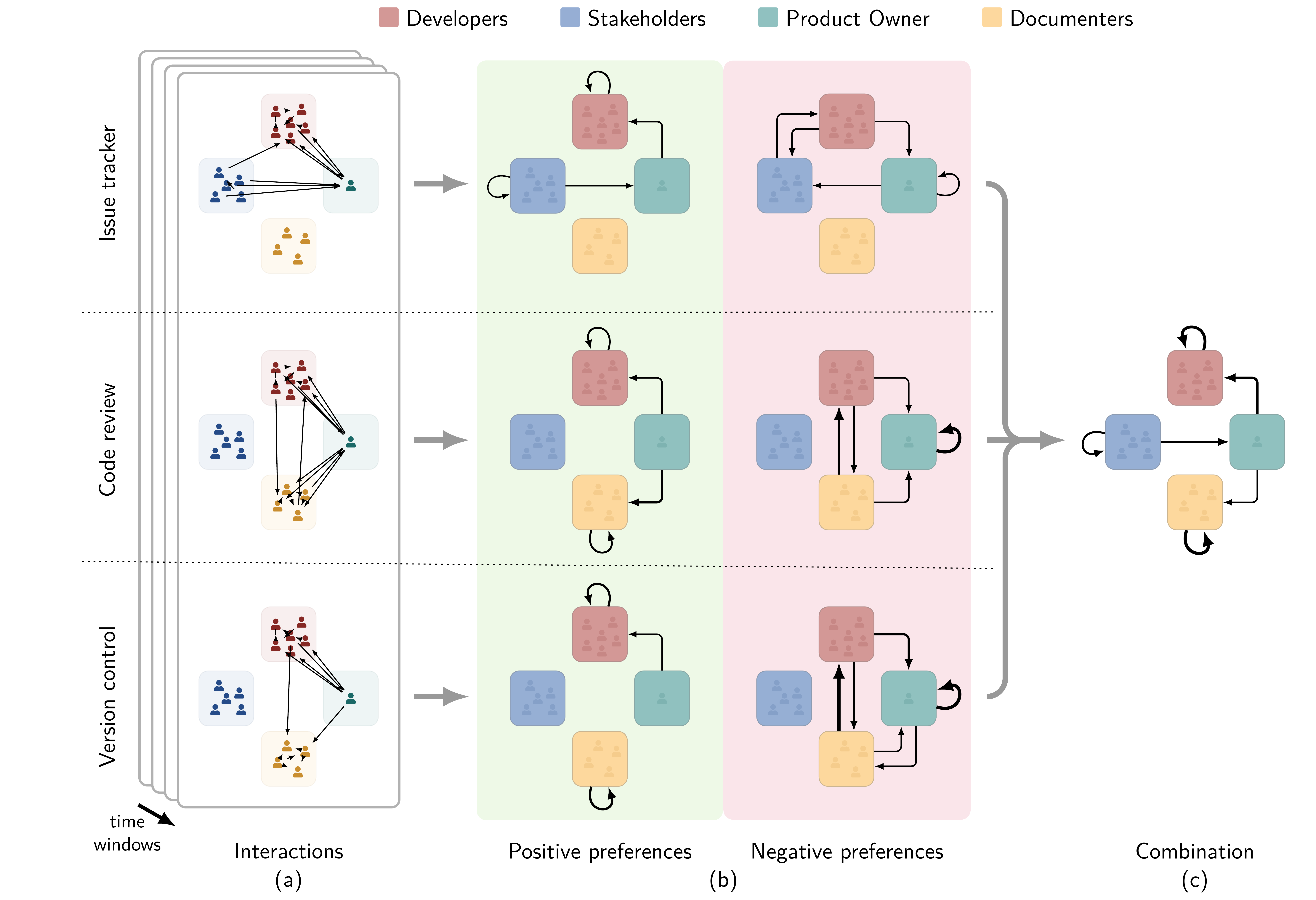}
    \caption{
        Applying our block model approach to extract the functional interaction structure for the studied project team at \genua{}.
        (a) Interactions between team members represented as interaction networks for each platform and year between 2010 and 2018.
        (b) Interaction preferences computed separately for positive ($>25\%$) and negative ($<25\%$) preferences and averaged across years.
        The width of links corresponds to the strength of the positive or negative preference.
        (c) The functional interaction structure of the team obtained as the combination of the positive preferences across the three platforms.
        By controlling for the activity of the team members, the block model approach allows for deeper insights going beyond the interaction counting approach, which only identified two categories of links.
    }\label{fig:bias-networks}
\end{figure*}

\paragraph{Negative interaction preferences}
We visualise these negative interaction preferences in the right column of \cref{fig:bias-networks}b.
The positive preferences discussed above imply the existence of corresponding negative preferences towards the other roles.
This means that the structure of the negative preferences, i.e., which links exist, is complementary to the structure of the positive preferences.
Therefore---rather than its structure---we are particularly interested in the \emph{strength} of the negative preferences displayed by each role.\\
\stakeholders{} are only active on the issue tracker, where they have positive interaction preferences with themselves and the \productowner{}.
As \documenters{} are not active on the issue tracker, \stakeholders{} only have a negative interaction preference towards \developers{}.\\
The \productowner{} is active on all platforms.
As indicated by the thin width of the links, the \productowner{} does not show strong negative interaction preferences to other roles.
However, we find a self-loop suggesting that \productowners{} interact significantly less than expected among themselves.
This is intuitive, as there is only one \productowner{} active at any point in time.
Therefore, if we observe more than one \productowner{} in one of our yearly snapshots, this indicates a transition between the two at some point during the year.
However, as they are active consecutively and not simultaneously, we find fewer interactions than  their activity suggests.\\
\developers{} are also active on all platforms.
As indicated by the similar width of all edges from \developers{} to all other roles, \developers{} do not show strong negative interaction preferences towards any role.\\
Finally, \documenters{} are active on the code review and version control platforms.
For \documenters{}, we find a strong negative interaction preference towards \developers{}.

So far, we have discussed interaction preferences for each platform separately.
We now combine them to obtain the functional interaction structure of the team across all platforms.
As the positive and negative interaction preferences are complementary, both reveal the same interaction structure.
However, positive interaction preferences are more natural to interpret.
Therefore, we show the functional interaction structure obtained by combining the positive interaction preferences in \cref{fig:bias-networks}c.
Overall, we find that \stakeholders{} represent the input into the development team.
\stakeholders{} interact primarily with the \productowner{} who, in turn, has strong interaction preferences towards \developers{} and \documenters{}.
These two roles represent sinks in the team, primarily interacting among their own role and not with each other.

\section{From Interaction Structure To Organigraph}\label{sec:interviews}

We now validate the detected interaction structure and enrich it with information on the \emph{function} of these relations---yielding the team's organigraph.
To this end, we conducted interviews with five experienced members of the studied product team (\code{Int1} -- \code{Int5}).
Our set of interviewees consists of two \developers{}, two \stakeholders{}, and one former \productowner{}, ensuring that we get a broad range of first-hand perspectives into the product team's development processes.
All interviews were held online in a video chat in March~2021 and lasted for approx.\ 60~minutes, followed by a debriefing.
We set up each interview in a semi-structured format, combining closed-ended survey questions with open-ended discussions \citep{Adams2015}.
The interviews were conducted without the aid of any additional material.
As we summarise in the following, the interviews validated and explained all interaction preferences identified in our quantitative analysis (cf. \cref{sec:quantitative_analysis}).
We show the organigraph enriching our quantitative results with the explanations from the interviews in \cref{fig:genua-organigraph}.

\begin{figure}
    \centering
    \includegraphics[width=.65\linewidth]{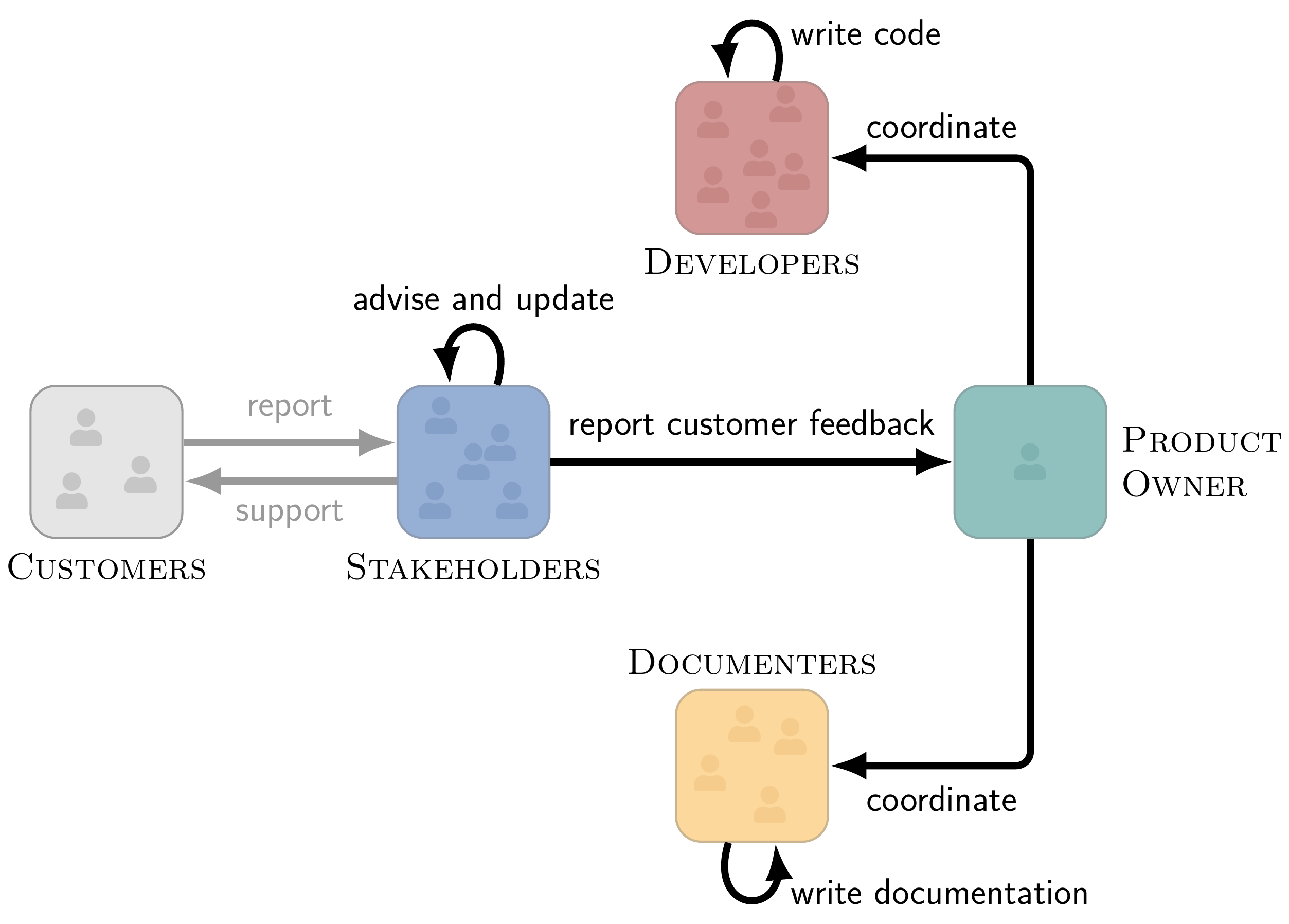}
    \caption{
        Organigraph of the product team at \genua{} across all three development platforms. To obtain the organigraph, we start from the functional interaction structure detected using our block model approach (see \cref{fig:bias-networks}c). We then enrich it with the information collected during our interviews to allow for an interpretation.
    }\label{fig:genua-organigraph}
\end{figure}

\subsection{The role of Stakeholders}

As we discussed in \cref{sec:roles}, the majority of \stakeholders{} are customer-facing and often located at the customer's sites.
Consequently, they are the first to learn about new bugs, required features, or new use cases for which they forward feedback directly to the \productowner{}.
\genuastatement{The customers' wishes for new features are supposed to be assigned to the \productowner{}.}{[Statement by \code{Int3}]}
\genuastatement{\stakeholders{} discuss new features always in direct coordination with the PO.}{[Statement by \code{Int4}]}
Simultaneously, the geographical distribution of \stakeholders{} explains their reduced interactions with \documenters{} and \developers{}.
\genuastatement{\stakeholders{} are not at the company's [i.e., \genua{}'s] site, and therefore can't just go into a \developer{}'s office and ask.}{[Statement by \code{Int4}]}
Further, \stakeholders{} do not have access to the code review and version control platforms, explaining the observed lack of interactions there.
\genuastatement{Actually, normal \stakeholders{} have nothing to do with \aegis{} [the code review platform] and \git{} [the version control system].}{[Statement by \code{Int3}]}
Internally, \stakeholders{} update and advise each other on common problems and critical bugs, which they then champion to be prioritised in the team's weekly bug meetings.

\subsection{The role of the Product Owner}
Collecting the information from \stakeholders{}, the \productowner{} leads the weekly bug meetings and is responsible for scheduling and prioritising what is being worked on.
\genuastatement{In bug meetings, the \productowner{}, some \stakeholders{}, and also a couple of \developers{}, who took care of the bugs, discuss prioritisation, and their initial analysis.}{[Statement by \code{Int1}]}
The \productowner{} then coordinates and oversees the rest of the team.
Thus, the \productowner{} indeed acts as a fixed mediator for feedback from the \stakeholders{} to the \developers{} and \documenters{}, confirming the results of our quantitative analysis.

\subsection{The roles of Developers and Documenters}

Following the bug meeting, the \developers{} work on changes resolving the bugs or implementing the discussed features and \documenters{} update the product's documentation accordingly.
\genuastatement{Based on the outcome of the bug meeting, the \developers{} develop. And the \documenters{} 
can, of course, also see what is written and then document this.}{[Statement by \code{Int1}]}
All interviewees agreed that these two processes occur mostly independently, explaining the infrequent interactions between \developers{} and \documenters{}.
\genuastatement{\documenters{} and \developers{} have their closed problem domains. The \developer{} tries to get a feature working from a technical perspective, and the \documenter{} tries to explain it to a user at the other end.}{[Statement by \code{Int5}]}
However, interactions within their own role still occur very frequently.
One key reason for this is \genua{}'s internal review process, which requires all changes---including changes to the documentation---to be developed, reviewed, and integrated by three different team members, which automatically sparks interactions between many different members.
\genuastatement{Whenever something is changed, someone has to look at it [i.e., review and integrate it into the codebase].}{[Statement by \code{Int5}]}

In conclusion, with our interviews, we could validate and explain all detected interaction preferences---both positive and negative---between roles.
This validates our quantitative approach and shows that we can extract the functional interaction structure of teams accurately and in a computationally efficient manner.

\section{Interaction Structure Optimisation}\label{sec:res-optimisation-genua}

In \cref{sec:quantitative_analysis}, we started our analysis of the product team from the observed individual interactions between team members.
We then grouped the team members according to their role in the team, yielding the functional interaction structure capturing the interaction preferences between members of different roles reported in \cref{fig:bias-networks}c.
Through this step, we aggregated all team members of a given role into a single representative node in the resulting role interaction network.
Implicitly, this assumes that all members of a role are similar to the degree that they can be considered as interchangeable.
This strong assumption is unlikely to be fully fulfilled in any real-world organisation (cf. \cref{sec:related-work} for a discussion).
The role definitions from \cref{sec:roles} already state that, e.g., \stakeholders{} take on multiple different functions and specialisations.
Similarly, our interviews also suggest a degree of heterogeneity amongst \developers{}, both in terms of their experience and their knowledge of different parts of the codebase.
\genuastatement{I think there are still comfort zones where people make initial changes and whom you let do it [make changes in a specific area of the codebase].}{[Statement by \code{Int5}]}
However, by employing agile development methods, the team actively aims to promote and enhance knowledge diffusion among the \developers{}, to reduce the risk of knowledge loss when a member leaves the team.
\genuastatement{One of the philosophies of Scrum is that `everyone can do everything' to address precisely the problems arising when the bus comes [referring to the truck factor, which is also known as the bus factor], or Google simply pays more. Thus we try to counteract exactly these problems in advance through XP [Extreme Programming] and pair programming [deliberate pairing of team members with different expertise].}{[Statement by \code{Int1}]}
This motivates a final experiment in which we use our extracted interaction preferences to assess where the team currently stands and to what extent further homogeneity among the members of roles could improve knowledge diffusion within the team.

To quantify knowledge diffusion, we use the measure \emph{potentiality} ($\text{Pot}$) proposed in \citep{Zingg2019}.
Potentiality utilises the notion of entropy to quantify the extent to which members distribute their interactions across the entire team rather than among a few specific collaborators:
\begin{equation}
    \text{Pot} := \frac{H^\text{observed}}{H^\text{max}} \in [0, 1].
\end{equation}
Here, $H^\text{observed}$ is the entropy of the observed interaction distribution, and $H^\text{max}$ is the highest possible entropy achieved when all team members interact with everyone else equally often.
A potentiality close to $1$ indicates that most members interact with the entire team, whereas a potentiality close to $0$ indicates that many members only have a few interaction partners.

As we discussed in \cref{sec:bccm-approach}, we encode the team's interaction \emph{structure} through the $\omega$ parameters of the corresponding BCCM (cf. \cref{eq:bccm-omega}).
In contrast, potentiality is computed on interaction \emph{networks}.
Given an interaction structure---i.e., a specified BCCM---we obtain the distribution of likely interaction networks using the sampling approach implemented in \citep{ghypernet}.
Subsequently, we compute potentiality for all sampled networks obtaining a distribution of values capturing---and hence allowing us to compare---the team's knowledge diffusion for different interaction structures.

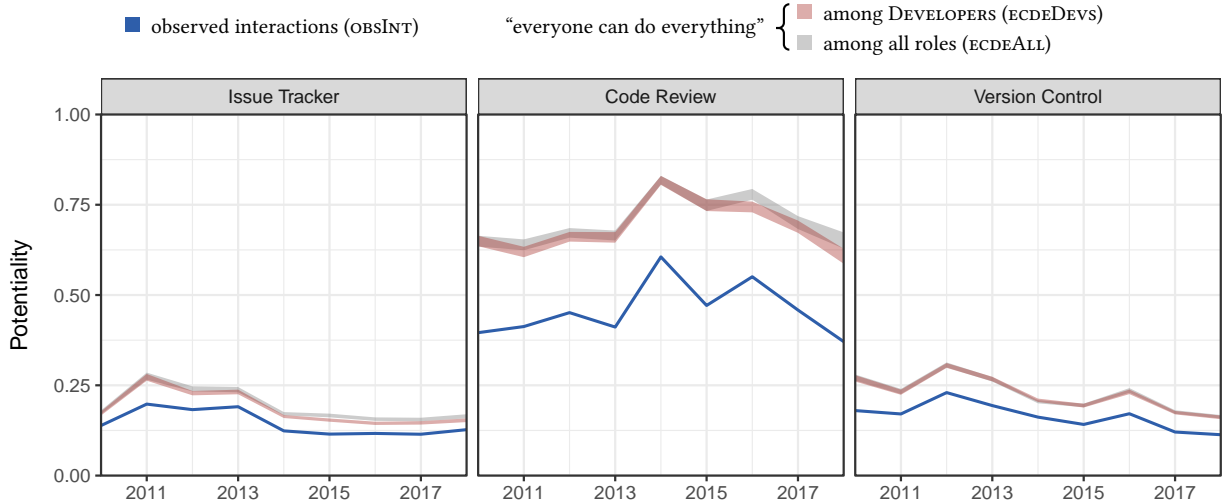
\begin{figure*}
    \centering
    \begin{tikzpicture} \footnotesize
        \node (Plot) at (0, 0) {
            \includegraphics[width=\textwidth, trim=5 28 7 8, clip]{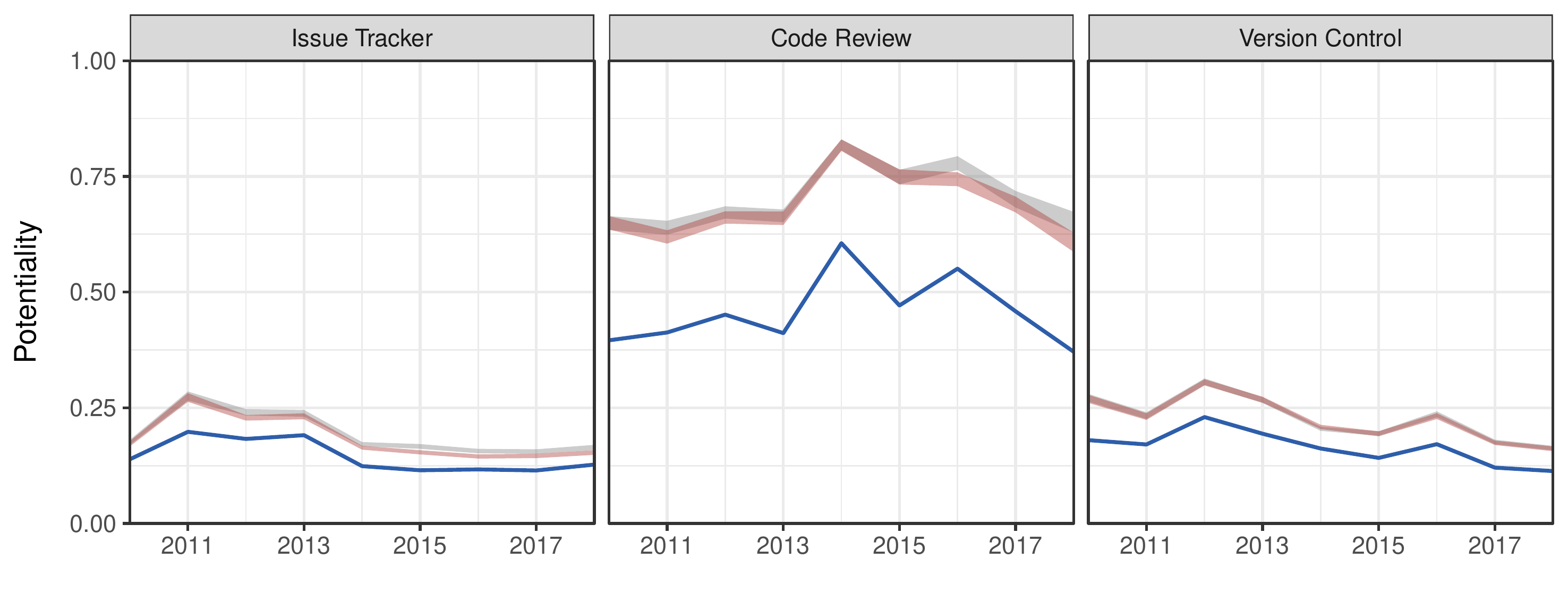}
        };
        \node[above=0.4em of Plot] {
            \begin{tikzpicture}
                \node[minimum width=0.5em, fill=colorScenarioObserved] (RectValueRange) at (0, 0) {};
                \node[right=0.5em of RectValueRange.north east, anchor=north west, inner sep=0em] (TextValueRange) {observed interactions (\code{obsInt})};

                \node[right=4em of TextValueRange.north east, anchor=north west, inner sep=0em] (TextEcdE) {``everyone can do everything''};

                \node[right=1.5em of TextEcdE.north east, anchor=north west, minimum width=0.5em, fill=colorScenarioAgile, yshift=1.7mm] (RectEmpirical) {};
                \node[right=0.5em of RectEmpirical.north east, anchor=north west, inner sep=0em] (TextEmpirical) {among \developers{} (\code{ecdeDevs})};

                \node[right=1.5em of TextEcdE.north east, anchor=north west, minimum width=0.5em, fill=colorScenarioRoles, yshift=-2.3mm] (RectDevOnly) {};
                \node[right=0.5em of RectDevOnly.north east, anchor=north west, inner sep=0em] (TextDevOnly) {among all roles (\code{ecdeAll})};

                \draw[thick, rounded corners=2] ($(RectEmpirical.north west) + (-1mm, 0)$) -| ($(TextEcdE.east) + (2.5mm, 0)$) -- ($(TextEcdE.east) + (1mm, 0)$) -- ($(TextEcdE.east) + (2.5mm, 0)$) |- ($(RectDevOnly.south west) + (-1mm, 0)$);
            \end{tikzpicture}
        };
    \end{tikzpicture}
    \caption{
        Potentiality as a measure for knowledge diffusion. We compare the observed interactions (\code{obsInt}) against two synthetical benchmark scenarios proposed by \genua{}. The first scenario (\code{ecdeDevs}) assumes homogeneity among all \developers{}. The second scenario (\code{ecdeAll}) assumes homogeneity among members of all roles.
        We show the results for all years and development platforms separately.
        The team shows knowledge diffusion in the upper third of the attainable range defined by the benchmark scenarios.
        The two benchmark scenarios result in almost identical levels of knowledge diffusion.
    }\label{fig:change-potentials}
\end{figure*}

We report our results in \cref{fig:change-potentials}.
In \textcolor{colorScenarioObserved}{\faSquare}, we show the potentiality computed for the observed interactions (\code{obsInt}) over time.
We compare the knowledge diffusion in the observed case to two synthetically created benchmark scenarios suggested by \genua{}.
In the first scenario (\code{ecdeDevs}), shown in \textcolor{colorScenarioAgile}{\faSquare}, we assume that the team achieves the stated aim that ``everyone can do everything'' (\code{ecde}) among \developers{}, effectively making them interchangeable.
This corresponds to a BCCM model where all developers are aggregated into a single block, while all other team members are represented by individual blocks.
Finally, in the second scenario (\code{ecdeAll}) shown in \textcolor{colorScenarioRoles}{\faSquare}, we assume that ``everyone can do everything'' holds not only for \developers{} but for all roles.
This corresponds to a BCCM where, analogous to the organigraph in \cref{fig:genua-organigraph}, all team members are aggregated into four blocks corresponding to their role.

For all platforms and all years we observe $\text{Pot}(\code{ecdeAll}) \geq \text{Pot}(\code{ecdeDevs}) \geq \text{Pot}(\code{obsInt})$.
This ordering aligns with our expectation that knowledge diffusion has an inverse relation to the heterogeneity of members of a role.

Notably, the difference between \code{ecdeDevs} and \code{ecdeAll} is always diminishingly small.
This means that almost all possible improvements in knowledge diffusion can already be achieved if ``everyone can do everything'' among \developers{}.
As indicated by \code{obsInt}, the team currently achieves a knowledge diffusion corresponding to around 70\% of the optimal case \code{ecdeAll}.
Our analysis suggests that to improve this further, the team should target knowledge diffusion among \developers{} first.

Comparing the three platforms, we observe significantly higher knowledge diffusion on the code review platform.
This indicates that \genua{}'s efforts to promote interactions by requiring that at least three different team members contribute to all changes are successful.
Finally, the code review platform is primarily used by \developers{}, explaining why the difference between \code{obsInt} and \code{ecdeDevs} is largest here.

In conclusion, we find that the studied team achieves knowledge diffusion in the upper third of the attainable range.
Our analysis shows that almost the entire remaining gain can already be achieved by obtaining optimal knowledge diffusion among developers (cf. \code{ecdeDevs}).
Working towards this, the team at \genua{} implements various agile methods, including Scrum, Extreme Programming, and pair programming.

\section{Threats to Validity}\label{sec:threats-to-validity}

Our study is subject to some threats to validity which we discuss in the following.

\paragraph{Internal validity}
While we have taken the utmost effort and care to obtain complete and correct data on the interactions among all members of the product team, there are three limitations that we discuss in the following.\\
First, for our study, we mined all actions logged in the complete databases of all three development platforms used by the analysed team.
From our discussions with team members, we learned that no development occurs without generating entries on these platforms as the team strictly enforces all bugs and feature requests to be tracked and version controlled.
That said, due to confidentiality concerns, we could not obtain and analyse any text data.
Next to the content of the interactions on the three development platforms, this also means that we did not have access to any email or chat communication.
Finally, interactions such as personal discussions are not recorded.
As a consequence, these interactions are missing from our data.\\
Second, as discussed in \cref{sec:interactions}, the development platforms record actions instead of interactions between team members.
In discussion with members of the product team, we identified a set of practices allowing us to extract the interactions corresponding to the recorded actions.
However, we expect a degree of heterogeneity in the behaviour of team members, which is not covered by the practices.
Furthermore, we expect team members to adapt their behaviour over time and based on the context of the situation, resulting in changes over time.\\
Finally, the three development platforms record different types of interactions (discussions, code reviews, and co-editing of code).
In our interviews, we discussed the possibility of weighing the different types of interactions for our combined results.
However, there was no consensus among our interviewees concerning which platform is most important for them with regard to knowledge exchange.
Therefore, for our combined results, we treated all platforms as equal.

\paragraph{Construct validity}
In the first part of our study, we aimed to extract the functional relations between roles in a product team.
To this end, we studied the team's interactions using a block model approach.
The resulting relations match those identified in our subsequent interviews, confirming the usefulness of our approach.
However, we cannot entirely rule out unlikely cases in which our approach missed relations that none of the five interviewed team members was aware of, as those would show up neither in our data nor the interviews.\\
In the second part, we used the resulting functional interaction structure to assess knowledge diffusion in the team.
To this end, we used the entropy-based measure \textit{potentiality}.
While our results suggest that potentiality captures knowledge diffusion adequately, additional measures, e.g., also capturing the content of interactions, could further improve our analysis.
Unfortunately, as we did not have access to any text data, we could not further explore this.\\
Finally, for our two hypothetical benchmark scenarios \code{ecdeDevs} and \code{ecdeAll}, we assumed perfect homogeneity among all team members of a role.
We argue that the scenarios are helpful as optimal cases the team can work towards.
However, different activity levels of team members, turnover, and differences in team members' experience with the product will always cause the scenarios to remain purely theoretical.
In addition, our analysis excluded the discussion of the benefits of heterogeneity, e.g., increased productivity and creativity \citep{shin2007educational,hamilton2003team}, which we will assess in future work.

\paragraph{External validity}
Lastly, we performed our analysis in a case study for a single product team at \genua{}, which sparks questions regarding the external validity of our analysis.
Our analysis approach solely relies on time-stamped interaction data and information regarding team members' roles and makes no assumptions concerning their content.
Therefore, we do not see concerns regarding the generalisability of our approach.

\section{Related Work}\label{sec:related-work}

In this paper, we introduced an approach to measure the functional interaction structure of a team.
We also exemplified how our approach can be used to reveal community smells, i.e., flaws in the interaction structure.
Finally, we compared the interaction structure against synthetic benchmark scenarios generating insights on how it can be improved.
We discuss the related work for these three aspects separately.

\paragraph{Interaction structure detection} The interaction structure in teams has been studied and characterised in a broad range of empirical studies.
Commonly, this is achieved via a network approach.
Here, researchers represent individuals as nodes and their interactions as edges.
They then compute various network measures to characterise their interaction structure \citep{Vijayaraghavan2015a,Zingg2020,Casiraghi2020}.
For example, the betweenness centrality could reveal hubs in OSS teams who route the information flow from peripheral developers into the core team \citep{Valverde2007}.
Entropy measures could show that humans interact with a broad range of peers in the early stages of group formation but narrow down their contacts as time proceeds \citep{Kulisiewicz2018}.
Similarly, the \emph{potentiality}, an entropy-based measure, was used to quantify the distribution of interactions across a team \citep{Zingg2019}, thereby proxying the resilience to forming knowledge islands.
Using the degree assortativity and clustering coefficient, the impact of the departure of a core developer on an Open Source team was measured \citep{Zanetti2013}.
In a similar approach, the authors of \citep{Bettenburg2010} studied how well various network measures predict the risk of introducing software defects.
Using non-network approaches, the authors of \citep{Sailer2016} characterised the interaction structure spatially by detecting locations in an office building where employees frequently interact.
In other works, the task redistribution between software developers was studied with agent-based models \citep{Malarz2016,Casiraghi2021}.
The key difference to this work is that all approaches mentioned above study the interaction structure between \emph{individuals}.
Instead, we focus on the functional interaction structure derived from the interactions between members of different \emph{roles} in a team.

\paragraph{Community smells} Flaws in the interaction structure of software development teams are often referred to as \emph{community smells}.
For a thorough review, we refer to \citep{caballero2022community}.
Typically, community smells cause inefficiency by hindering the information exchange within a team.
As collections of knowledge workers, software development teams require and encourage the creation of specialised knowledge \citep{Faraj2000} such as technical knowledge on the implementation of specific code areas \citep{Zazworka2010} or long-term experience in a project \citep{Egeland2017}.
This can lead to the emergence of small sub-groups of developers that solely possess specific knowledge---so-called organisational silos.
If these structures are not detected and countered in a timely manner, they can put the team at risk, as this knowledge is lost if these members leave.
Thus, community smells can even threaten an entire team's existence \citep{Avelino2016}.
The authors of \citep{Sedano2017} show that regaining lost knowledge can require months of sifting through old code, commit logs, etc., in which the team makes little to no progress.
The risk behind organisational silos is amplified by the fast employee turnover in IT companies, which is typically below two years \citep{Peterson2017}.
The literature has proposed different measures to estimate the size of such organisational silos, e.g., the \emph{bus factor} or \emph{truck number} \citep{Avelino2016,Ferreira2019,Jabrayilzade2022}.
These works, however, focus on individuals and not on the group structure of a team.
Our approach enables the detection of community smells affecting the group structure as given by the members' roles.
Thereby, our approach can, for example, be used to identify roles with low knowledge diffusion, whose members bear the risk of forming organisational silos.

\paragraph{Interaction structure optimisation} Finally, the literature shows that teams can eliminate community smells effectively \emph{if} they are able to find them \citep[see, e.g.,][]{Catolino2020}.
Knowledge in software development teams is largely shared through interactions \citep{Rus2002}.
Therefore, community smells are typically countered by incentivising and increasing the number of interactions within the team to improve the distribution of knowledge \citep{Catolino2021}.
To distribute interactions, agile development frameworks encourage periodic team meetings \citep{Highsmith2001}, pair-programming \citep{Cockburn2001}, where two developers sit together at one computer and program, or pair-rotation \citep{Zazworka2010}, where the pairs are reshuffled periodically.
Other approaches pursuing similar aims are Scrum~\citep{Schwaber2020}, Kanban~\citep{Ahmad2013}, Extreme Programming~\citep{Beck1999}, or DevOps~\citep{Zhou2022}.
These approaches have become popular in modern IT teams, but judging how effective they are in a given team is challenging.
Closing this gap, our approach enables teams to assess and manage their functional interaction structure in real time.
In addition, by enabling teams to evaluate their current interaction structure against alternative benchmark scenarios, our approach facilitates the identification of optimal interaction structures the team can work towards.

\section{Conclusions}\label{sec:conclusions}

An efficient interaction structure facilitates knowledge diffusion, allowing the team to maintain its performance and retain its knowledge base even when team members leave.
However, particularly in flexible, self-organised teams, the interaction structure evolves over time.
While this allows the team to adapt to new challenges, it also bears the risk of undesirable outcomes, such as reduced software quality or the emergence of community smells.

This paper showed how the functional interaction structure, i.e., the preferences with which members of different roles interact with each other, can be directly inferred from the traces that team members create on their development platforms during their daily work.
To this end, we first demonstrated that approaches merely counting the interactions between members belonging to different roles are insufficient, as they fail to account for the unique activity levels of the team members.
Instead, we proposed an approach that considers how many interactions members of each role can initiate and receive using a block-constrained configuration model.
This allows us to compute the interaction preferences between members of different roles.

In a case study, we mined comprehensive data tracking the development process of a product team at the \genuaDescription{} \genuaLong{} across three development platforms.
We then applied our approach to extract the functional interaction structure of the team.

We conducted semi-structured interviews with five team members in which we validated the accuracy of the detected interaction structure.
In addition, the interviews allowed us to enrich the detected interaction structure with information on the purpose of each interaction.
This made the interaction structure interpretable and yielded the team's organigraph.

During the interviews, we further learned that to prevent knowledge loss, \genua{} strives for homogeneity among members of a role---i.e., ``everyone can do everything'' across members of a role.
This motivated a final experiment in which we showed how our approach enables teams to compare themselves against synthetic benchmark scenarios.
Specifically, we studied the knowledge diffusion in the development team and compared it to two scenarios suggested by \genua{}.
The first scenario assumed homogeneity only among \developers{}, while the second scenario assumed homogeneity for all roles.
Our results demonstrated that the team currently reaches knowledge diffusion in the upper third of the attainable range.
We further showed that reaching homogeneity for all roles in the team is not required.
Instead, almost all possible gains in knowledge diffusion can already be achieved by further promoting interactions between \developers{}, which the team does by applying Scrum, Extreme Programming, and pair programming.
Our approach is computationally efficient, allowing the team to track the results of their efforts and manage their interaction structure in real-time, based solely on readily available development data.

\section*{Acknowledgements}
We thank the 5 anonymous interviewees from \genua{} for the valuable insights provided during the interviews.
We further thank Giona Casiraghi for useful discussions about the BCCM and contributions to the R implementation of the Potentiality function.

\bibliographystyle{sg-bibstyle}
\bibliography{bibliography}

\begin{thebibliography}{54}
\expandafter\ifx\csname natexlab\endcsname\relax\def\natexlab#1{#1}\fi
\expandafter\ifx\csname url\endcsname\relax
  \def\url#1{\texttt{#1}}\fi
\expandafter\ifx\csname urlprefix\endcsname\relax\def\urlprefix{URL }\fi
\expandafter\ifx\csname selectlanguage\endcsname\relax
  \def\selectlanguage#1{\relax}\fi

\bibitem[{Adams(2015)}]{Adams2015}
Adams, W.~C. (2015).
\newblock Conducting semi-structured interviews.
\newblock In: \emph{Handbook of Practical Program Evaluation}, Hoboken, NJ,
  USA: John Wiley {\&} Sons, Inc. pp. 492--505.

\bibitem[{Ahmad \emph{et~al.}(2013)Ahmad, Markkula and Oivo}]{Ahmad2013}
Ahmad, M.~O.; Markkula, J.; Oivo, M. (2013).
\newblock {Kanban in software development: A systematic literature review}.
\newblock In: \emph{2013 39th Euromicro Conference on Software Engineering and
  Advanced Applications}. IEEE, pp. 9--16.

\bibitem[{Avelino \emph{et~al.}(2016)Avelino, Passos, Hora and
  Valente}]{Avelino2016}
Avelino, G.; Passos, L.; Hora, A.; Valente, M.~T. (2016).
\newblock {A novel approach for estimating truck factors}.
\newblock In: \emph{2016 IEEE 24th International Conference on Program
  Comprehension (ICPC)}. IEEE, vol. 2016-July, pp. 1--10.

\bibitem[{Beck(1999)}]{Beck1999}
Beck, K. (1999).
\newblock \emph{{Extreme Programming explained: Embrace change}}.
\newblock USA: Addison-Wesley Longman Publishing Co., Inc.

\bibitem[{Beck \emph{et~al.}(2001)Beck, Beedle, Van~Bennekum, Cockburn,
  Cunningham, Fowler, Grenning, Highsmith, Hunt, Jeffries
  \emph{et~al.}}]{beck2001manifesto}
Beck, K.; Beedle, M.; Van~Bennekum, A.; Cockburn, A.; Cunningham, W.; Fowler,
  M.; Grenning, J.; Highsmith, J.; Hunt, A.; Jeffries, R.; \emph{et~al.}
  (2001).
\newblock Manifesto for agile software development.
\newblock \emph{Agile Alliance} .

\bibitem[{Bettenburg and Hassan(2010)}]{Bettenburg2010}
Bettenburg, N.; Hassan, A.~E. (2010).
\newblock Studying the Impact of Social Structures on Software Quality.
\newblock In: \emph{2010 IEEE 18th International Conference on Program
  Comprehension}. IEEE, pp. 124--133.

\bibitem[{Bolici \emph{et~al.}(2016)Bolici, Howison and Crowston}]{Bolici2016}
Bolici, F.; Howison, J.; Crowston, K. (2016).
\newblock {Stigmergic coordination in FLOSS development teams: Integrating
  explicit and implicit mechanisms}.
\newblock \emph{Cognitive Systems Research} \textbf{38}, 14--22.

\bibitem[{Caballero-Espinosa \emph{et~al.}(2022)Caballero-Espinosa, Carver and
  Stowers}]{caballero2022community}
Caballero-Espinosa, E.; Carver, J.~C.; Stowers, K. (2022).
\newblock Community smells—The sources of social debt: A systematic
  literature review.
\newblock \emph{Information and Software Technology} , 107078.

\bibitem[{Casiraghi(2019)}]{Casiraghi2019c}
Casiraghi, G. (2019).
\newblock {The block-constrained configuration model}.
\newblock \emph{Applied Network Science} \textbf{4(1)}, 123.

\bibitem[{Casiraghi and Nanumyan(2020)}]{ghypernet}
Casiraghi, G.; Nanumyan, V. (2020).
\newblock r-ghypernet Version 1.0.1.
\newblock \url{https://doi.org/10.5281/zenodo.4117523}.

\bibitem[{Casiraghi and Schweitzer(2020)}]{Casiraghi2020}
Casiraghi, G.; Schweitzer, F. (2020).
\newblock Improving the Robustness of Online Social Networks: A Simulation
  Approach of Network Interventions.
\newblock \emph{Frontiers in Robotics and AI} \textbf{7}, 57.

\bibitem[{Casiraghi \emph{et~al.}(2021)Casiraghi, Zingg and
  Schweitzer}]{Casiraghi2021}
Casiraghi, G.; Zingg, C.; Schweitzer, F. (2021).
\newblock {The downside of heterogeneity: How established relations counteract
  systemic adaptivity in tasks assignments}.
\newblock \emph{Entropy} \textbf{23(12)}, 1677.

\bibitem[{Cataldo \emph{et~al.}(2006)Cataldo, Wagstrom, Herbsleb and
  Carley}]{cataldo2006identification}
Cataldo, M.; Wagstrom, P.~A.; Herbsleb, J.~D.; Carley, K.~M. (2006).
\newblock Identification of coordination requirements: Implications for the
  design of collaboration and awareness tools.
\newblock In: \emph{Proceedings of the 2006 20th Anniversary Conference on
  Computer Supported Cooperative Work}. pp. 353--362.

\bibitem[{Catolino \emph{et~al.}(2021)Catolino, Palomba, Tamburri and
  Serebrenik}]{Catolino2021}
Catolino, G.; Palomba, F.; Tamburri, D.~A.; Serebrenik, A. (2021).
\newblock Understanding Community Smells Variability: A Statistical Approach.
\newblock In: \emph{2021 IEEE/ACM 43rd International Conference on Software
  Engineering: Software Engineering in Society (ICSE-SEIS)}. IEEE, vol.
  2021-May, pp. 77--86.
\newblock ISSN 02705257.

\bibitem[{Catolino \emph{et~al.}(2020)Catolino, Palomba, Tamburri, Serebrenik
  and Ferrucci}]{Catolino2020}
Catolino, G.; Palomba, F.; Tamburri, D.~A.; Serebrenik, A.; Ferrucci, F.
  (2020).
\newblock Refactoring community smells in the wild: The practitioner's field
  manual.
\newblock \emph{Proceedings - International Conference on Software Engineering}
  , 25--34.

\bibitem[{Cockburn and Williams(2001)}]{Cockburn2001}
Cockburn, A.; Williams, L. (2001).
\newblock {The costs and benefits of pair programming}.
\newblock In: \emph{Extreme Programming Examined}, USA: Addison-Wesley Longman
  Publishing Co., Inc. pp. 223--243.

\bibitem[{Egeland(2017)}]{Egeland2017}
Egeland, I. (2017).
\newblock \emph{Knowledge retention in organizations}.
\newblock Msc. thesis, University of Stavanger.

\bibitem[{Faraj and Sproull(2000)}]{Faraj2000}
Faraj, S.; Sproull, L. (2000).
\newblock {Coordinating expertise in software development teams}.
\newblock \emph{Management Science} \textbf{46(12)}, 1554--1568.

\bibitem[{Ferreira \emph{et~al.}(2019)Ferreira, Mombach, Valente and
  Ferreira}]{Ferreira2019}
Ferreira, M.; Mombach, T.; Valente, M.~T.; Ferreira, K. (2019).
\newblock {Algorithms for estimating truck factors: A comparative study}.
\newblock \emph{Software Quality Journal} \textbf{27(4)}, 1583--1617.

\bibitem[{Gote \emph{et~al.}(2022)Gote, Mavrodiev, Schweitzer and
  Scholtes}]{gote2022big}
Gote, C.; Mavrodiev, P.; Schweitzer, F.; Scholtes, I. (2022).
\newblock {Big data = Big insights? Operationalising Brooks' law in a massive
  GitHub data set}.
\newblock In: \emph{2022 IEEE/ACM 44th International Conference on Software
  Engineering (ICSE)}. pp. 262--273.

\bibitem[{Gote \emph{et~al.}(2019)Gote, Scholtes and Schweitzer}]{Gote2019}
Gote, C.; Scholtes, I.; Schweitzer, F. (2019).
\newblock {git2net - Mining time-stamped co-editing networks from large git
  repositories}.
\newblock In: \emph{2019 IEEE/ACM 16th International Conference on Mining
  Software Repositories (MSR)}. IEEE, vol. 2019-May, pp. 433--444.
\newblock ISSN 21601860.

\bibitem[{Gote \emph{et~al.}(2021)Gote, Scholtes and
  Schweitzer}]{gote2021analysing}
Gote, C.; Scholtes, I.; Schweitzer, F. (2021).
\newblock Analysing Time-Stamped Co-Editing Networks in Software Development
  Teams using git2net.
\newblock \emph{Empirical Software Engineering} \textbf{26(4)}, 1--41.

\bibitem[{Hamilton \emph{et~al.}(2003)Hamilton, Nickerson and
  Owan}]{hamilton2003team}
Hamilton, B.~H.; Nickerson, J.~A.; Owan, H. (2003).
\newblock Team incentives and worker heterogeneity: An empirical analysis of
  the impact of teams on productivity and participation.
\newblock \emph{Journal of Political Economy} \textbf{111(3)}, 465--497.

\bibitem[{Highsmith and Cockburn(2001)}]{Highsmith2001}
Highsmith, J.; Cockburn, A. (2001).
\newblock {Agile software development: The business of innovation}.
\newblock \emph{Computer} \textbf{34(9)}, 120--127.

\bibitem[{Hoegl and Gemuenden(2001)}]{Hoegl2001}
Hoegl, M.; Gemuenden, H.~G. (2001).
\newblock {Teamwork quality and the success of innovative projects: {A}
  theoretical concept and empirical evidence}.
\newblock \emph{Organization Science} \textbf{12(4)}, 435--449.

\bibitem[{Jabrayilzade \emph{et~al.}(2022)Jabrayilzade, Evtikhiev, Tuzun and
  Kovalenko}]{Jabrayilzade2022}
Jabrayilzade, E.; Evtikhiev, M.; Tuzun, E.; Kovalenko, V. (2022).
\newblock Bus Factor in Practice.
\newblock In: \emph{2022 IEEE/ACM 44th International Conference on Software
  Engineering: Software Engineering in Practice (ICSE-SEIP)}. IEEE, pp.
  97--106.
\newblock ISSN 02705257.

\bibitem[{Jassowski(2012)}]{jassowski2012organizational}
Jassowski, M. (2012).
\newblock Organizational dynamics: Understanding the impact of organizational
  structure in team productivity.
\newblock \emph{IEEE Design \& Test of Computers} \textbf{29(3)}, 52--59.

\bibitem[{Kulisiewicz \emph{et~al.}(2018)Kulisiewicz, Kazienko, Szymanski and
  Michalski}]{Kulisiewicz2018}
Kulisiewicz, M.; Kazienko, P.; Szymanski, B.~K.; Michalski, R. (2018).
\newblock {Entropy measures of human communication dynamics}.
\newblock \emph{Scientific Reports} \textbf{8(1)}, 15697.

\bibitem[{Lesyuk(2016)}]{Lesyuk2016}
Lesyuk, A. (2016).
\newblock \emph{{Mastering Redmine: An expert's guide to open source project
  management with Redmine}}.
\newblock Community Experience Distilled, Birmingham: Packt Publishing, 2nd
  edn.

\bibitem[{Malarz \emph{et~al.}(2016)Malarz, Kowalska-Styczeń and
  Kułakowski}]{Malarz2016}
Malarz, K.; Kowalska-Styczeń, A.; Kułakowski, K. (2016).
\newblock The working group performance modeled by a bi-layer cellular
  automaton.
\newblock \emph{Simulation} \textbf{92}, 179--193.

\bibitem[{Miller(2013)}]{Miller2013}
Miller, P. (2013).
\newblock {Aegis}.
\newblock \url{https://sourceforge.net/projects/aegis/}.

\bibitem[{Mintzberg and der Heyden(1999)}]{Mintzberg1999}
Mintzberg, H.; der Heyden, L.~V. (1999).
\newblock Organigraphs: Drawing how companies really work.
\newblock \emph{Harvard Business Review} \textbf{77}, 87--94, 184.

\bibitem[{Nakakoji \emph{et~al.}(2002)Nakakoji, Yamamoto, Nishinaka, Kishida
  and Ye}]{Nakakoji2002}
Nakakoji, K.; Yamamoto, Y.; Nishinaka, Y.; Kishida, K.; Ye, Y. (2002).
\newblock {Evolution patterns of open-source software systems and communities}.
\newblock In: \emph{Proceedings of the International Workshop on Principles of
  Software Evolution - IWPSE '02}. New York, New York, USA: ACM Press, p.~76.

\bibitem[{Palomba \emph{et~al.}(2018)Palomba, Tamburri, Fontana, Oliveto,
  Zaidman and Serebrenik}]{palomba2018beyond}
Palomba, F.; Tamburri, D.~A.; Fontana, F.~A.; Oliveto, R.; Zaidman, A.;
  Serebrenik, A. (2018).
\newblock Beyond technical aspects: How do community smells influence the
  intensity of code smells?
\newblock \emph{IEEE Transactions on Software Engineering} \textbf{47(1)},
  108--129.

\bibitem[{Peterson(2017)}]{Peterson2017}
Peterson, B. (2017).
\newblock Employee Retention Rate at Top Tech Companies.
\newblock
  \url{https://www.businessinsider.com/employee-retention-rate-top-tech-companies-2017-8}.

\bibitem[{Ralph \emph{et~al.}(2016)Ralph, Chiasson and Kelley}]{Ralph2016}
Ralph, P.; Chiasson, M.; Kelley, H. (2016).
\newblock Social theory for software engineering research.
\newblock In: \emph{Proceedings of the 20th International Conference on
  Evaluation and Assessment in Software Engineering}. ACM, pp. 1--11.

\bibitem[{Rank(2008)}]{Rank2008}
Rank, O.~N. (2008).
\newblock Formal structures and informal networks: Structural analysis in
  organizations.
\newblock \emph{Scandinavian Journal of Management} \textbf{24}, 145--161.

\bibitem[{Rus and Lindvall(2002)}]{Rus2002}
Rus, I.; Lindvall, M. (2002).
\newblock {Knowledge management in software engineering}.
\newblock \emph{IEEE Software} \textbf{19(3)}, 26--38.

\bibitem[{Sailer \emph{et~al.}(2016)Sailer, Koutsolampros, Austwick, Varoudis
  and Hudson-Smith}]{Sailer2016}
Sailer, K.; Koutsolampros, P.; Austwick, M.~Z.; Varoudis, T.; Hudson-Smith, A.
  (2016).
\newblock Measuring Interaction in Workplaces.
\newblock In: \emph{Architecture and Interaction: Human Computer Interaction in
  Space and Place}, Springer, Cham. pp. 137--161.

\bibitem[{Scholtes \emph{et~al.}(2016)Scholtes, Mavrodiev and
  Schweitzer}]{scholtes2016aristotle}
Scholtes, I.; Mavrodiev, P.; Schweitzer, F. (2016).
\newblock {From Aristotle to Ringelmann: A large-scale analysis of team
  productivity and coordination in Open Source Software projects}.
\newblock \emph{Empirical Software Engineering} \textbf{21(2)}, 642--683.

\bibitem[{Schwaber and Sutherland(2020)}]{Schwaber2020}
Schwaber, K.; Sutherland, J. (2020).
\newblock {The Scrum Guide - The definitive guide to Scrum: The rules of the
  game}.
\newblock \url{https://scrumguides.org/download.html}.

\bibitem[{Sedano \emph{et~al.}(2017)Sedano, Ralph and Peraire}]{Sedano2017}
Sedano, T.; Ralph, P.; Peraire, C. (2017).
\newblock Software Development Waste.
\newblock In: \emph{2017 IEEE/ACM 39th International Conference on Software
  Engineering (ICSE)}. IEEE, pp. 130--140.

\bibitem[{Shin and Zhou(2007)}]{shin2007educational}
Shin, S.~J.; Zhou, J. (2007).
\newblock {When is educational specialization heterogeneity related to
  creativity in research and development teams? Transformational leadership as
  a moderator}.
\newblock \emph{Journal of Applied Psychology} \textbf{92(6)}, 1709.

\bibitem[{Tamburri \emph{et~al.}(2016)Tamburri, Kazman and
  Fahimi}]{tamburri2016architect}
Tamburri, D.~A.; Kazman, R.; Fahimi, H. (2016).
\newblock The architect's role in community shepherding.
\newblock \emph{IEEE Software} \textbf{33(6)}, 70--79.

\bibitem[{Tamburri \emph{et~al.}(2015)Tamburri, Kruchten, Lago and van
  Vliet}]{Tamburri2015}
Tamburri, D.~A.; Kruchten, P.; Lago, P.; van Vliet, H. (2015).
\newblock Social debt in software engineering: Insights from industry.
\newblock \emph{Journal of Internet Services and Applications} \textbf{6}, 10.

\bibitem[{Tamburri \emph{et~al.}(2022)Tamburri, Kazman and
  Fahimi}]{tamburri2022relationship}
Tamburri, D. A.~A.; Kazman, R.; Fahimi, H. (2022).
\newblock On the Relationship Between Organisational Structure Patterns and
  Architecture in Agile Teams.
\newblock \emph{IEEE Transactions on Software Engineering} .

\bibitem[{Valverde and Solé(2007)}]{Valverde2007}
Valverde, S.; Solé, R.~V. (2007).
\newblock Self-organization versus hierarchy in open-source social networks.
\newblock \emph{Physical Review E} \textbf{76}, 046118.

\bibitem[{Vijayaraghavan \emph{et~al.}(2015)Vijayaraghavan, No{\"{e}}l, Maoz
  and D'Souza}]{Vijayaraghavan2015a}
Vijayaraghavan, V.~S.; No{\"{e}}l, P.-A.; Maoz, Z.; D'Souza, R.~M. (2015).
\newblock {Quantifying dynamical spillover in co-evolving multiplex networks}.
\newblock \emph{Scientific Reports} \textbf{5(1)}, 15142.

\bibitem[{Yang and Tang(2004)}]{yang2004team}
Yang, H.-L.; Tang, J.-H. (2004).
\newblock {Team structure and team performance in IS development: A social
  network perspective}.
\newblock \emph{Information \& management} \textbf{41(3)}, 335--349.

\bibitem[{Zanetti \emph{et~al.}(2013)Zanetti, Scholtes, Tessone and
  Schweitzer}]{Zanetti2013}
Zanetti, M.~S.; Scholtes, I.; Tessone, C.~J.; Schweitzer, F. (2013).
\newblock {The rise and fall of a central contributor: Dynamics of social
  organization and performance in the GENTOO community}.
\newblock In: \emph{2013 6th International Workshop on Cooperative and Human
  Aspects of Software Engineering, CHASE 2013 - Proceedings}. IEEE, pp. 49--56.

\bibitem[{Zazworka \emph{et~al.}(2010)Zazworka, Stapel, Knauss, Shull, Basili
  and Schneider}]{Zazworka2010}
Zazworka, N.; Stapel, K.; Knauss, E.; Shull, F.; Basili, V.~R.; Schneider, K.
  (2010).
\newblock Are developers complying with the process: An XP study.
\newblock In: \emph{Proceedings of the 2010 ACM-IEEE International Symposium on
  Empirical Software Engineering and Measurement - ESEM '10}. New York, New
  York, USA: ACM Press, pp. 1--10.

\bibitem[{Zhou \emph{et~al.}(2022)Zhou, Huang, Zhang, Huang, Shao and
  Zhong}]{Zhou2022}
Zhou, X.; Huang, H.; Zhang, H.; Huang, X.; Shao, D.; Zhong, C. (2022).
\newblock A Cross-Company Ethnographic Study on Software Teams for DevOps and
  Microservices: Organization, Benefits, and Issues.
\newblock In: \emph{Proceedings of the 44th International Conference on
  Software Engineering: Software Engineering in Practice (ICSE-SEIP)}. IEEE,
  pp. 1--10.

\bibitem[{Zingg \emph{et~al.}(2019)Zingg, Casiraghi, Vaccario and
  Schweitzer}]{Zingg2019}
Zingg, C.; Casiraghi, G.; Vaccario, G.; Schweitzer, F. (2019).
\newblock {What is the entropy of a social organization?}
\newblock \emph{Entropy} \textbf{21(9)}, 901.

\bibitem[{Zingg \emph{et~al.}(2020)Zingg, Nanumyan and Schweitzer}]{Zingg2020}
Zingg, C.; Nanumyan, V.; Schweitzer, F. (2020).
\newblock {Citations driven by social connections? A multi-layer representation
  of coauthorship networks}.
\newblock \emph{Quantitative Science Studies} \textbf{1(4)}, 1493--1509.

\end{thebibliography}

\end{document}